\documentclass[twocolumn,nofootinbib,showpacs,prd]{revtex4}
\usepackage{graphics,graphicx}
\usepackage{amsmath}
\usepackage{psfrag}
\usepackage{dcolumn}
\usepackage{color}

\allowdisplaybreaks

\def\no{\nonumber \\}

\newcommand{\be}{\begin{equation}}
\newcommand{\ee}{\end{equation}}
\newcommand{\bea}{\begin{eqnarray}}
\newcommand{\eea}{\end{eqnarray}}
\def\no{\nonumber \\}

\newcommand{\nn}{\nonumber}

\begin{document}

\title{
On the ability of various circular inspiral templates 
that incorporate radiation reaction effects
at the second post-Newtonian order
to capture 
inspiral gravitational waves from
compact binaries having tiny orbital eccentricities
}

\author{Manuel Tessmer}
\email{M.Tessmer@uni-jena.de}
\affiliation{Theoretisch-Physikalisches Institut,
Friedrich-Schiller-Universit\"at Jena,
Max-Wien-Platz 1,
07743 Jena, Germany}

\author{Achamveedu Gopakumar}
\email{A.Gopakumar@uni-jena.de}
\affiliation{Theoretisch-Physikalisches Institut,
Friedrich-Schiller-Universit\"at Jena,
Max-Wien-Platz 1,
07743 Jena, Germany}

\date{\today}

\begin{abstract}
 We probe the ability of various types of post-Newtonian(PN)-accurate circular
templates to capture
inspiral gravitational-wave (GW) signals from compact binaries having tiny orbital eccentricities.
The GW signals are constructed by adapting  
the phasing formalism, available in 
T.~Damour, A.~Gopakumar, and B.~R.~Iyer,
[Phys. Rev. D \textbf{70}, 064028 (2004)]. And, we employ the orbital energy 
and the time-eccentricity to describe the orbital evolution 
of eccentric binaries 
that incorporate the 
3PN-accurate conservative and the 2PN-accurate reactive dynamics.   
Using the fitting factor estimates, relevant for the initial LIGO, we show that  
circular templates, based on 
the adiabatic TaylorT1, complete adiabatic TaylorT1 and TaylorT4 approximants,
that require the 2PN-accurate energy flux are unable to capture our GW signals from
compact binaries having tiny residual orbital eccentricities.
However, the 2PN-order  
circular inspiral templates based on the recently introduced TaylorEt approximant
are found to be 
both {\it effectual} and {\it faithful} in capturing GWs from inspiralling compact binaries
having moderate eccentricities and we provide physical explanations for our 
observations.  We conclude that further investigations involving the actual 
interferometric data may be 
required to probe the ability of the widely employed 
traditional 2PN order circular templates
to capture inspiral GWs from astrophysical compact binaries, 
like the plausible compact binary progenitor candidate for the GRB~070201,
that are expected to contain tiny residual
orbital eccentricities. 

\end{abstract}


\maketitle

\section{Introduction}
  Inspiral GWs from non-spinning stellar mass compact binaries are being searched 
in the data generated by the operational 
first generation ground-based laser interferometric GW detectors 
~\cite{GWIFs}.
The GW data analysts employ the technique of {\it matched filtering} \cite{Helstrom}
that demands theoretically modeled inspiral templates to extract 
astrophysical GW signals from the noisy interferometric data.
The theoretical inspiral templates require 
the temporally evolving GW polarizations
$h_{+} (t)$ and $h_{\times}(t)$, 
and are usually constructed by
employing the PN
approximation to general relativity.
In the case of inspiralling compact binaries,
usually modeled to consist of point masses,
the  PN approximation provides, for example, the orbital dynamics
as corrections to the Newtonian  equations of motion
in terms of $({v}/{c})^2 \sim {G m}/{c^2\,r}$,
where $v$, $m$ and $r$ are
the characteristic orbital velocity,
the total mass, and the typical orbital separation,
respectively.

The LIGO Scientific Collaboration
(LSC) \cite{LSC} employs various types of PN-accurate inspiral templates, detailed in Ref.~\cite{DIS01},
for analyzing the LIGO data for inspiral GWs.
These templates are constructed by keeping only the Newtonian contributions to the 
amplitudes of $h_{+} (t)$ and $h_{\times}(t)$,
while their phase evolutions are
PN-accurate, resulting  
in the so-called restricted PN waveforms.    
While searching for inspiral GW signals from non-spinning compact binaries,
it turned out that 
the LSC had invoked various types of restricted 2PN order waveforms \cite{lsc_grb}.
These  inspiral template families provide 
slightly different prescriptions for the GW phase evolution that 
incorporate gravitational radiation reaction (RR) effects at the 2PN order, {\it i.e.}, 
$\left( {v}/{c} \right )^4 $ corrections beyond the Newtonian (quadrupolar) RR order.
These templates represent gravitational radiation field generated by
compact binaries 
inspiralling under the action of 2PN-accurate RR effects along circular orbits.
The approximation of quasi-circularity, namely inspiral along a sequence of circular orbits,
is justified due to the fact that the gravitational RR forces are highly 
efficient in  circularizing orbits of compact binaries \cite{P63}.

 Interestingly, the ability of the 2PN-accurate  circular templates,
available in Ref.~\cite{DIS01},
to detect GWs from  
compact binaries in inspiralling eccentric orbits is not explored in the 
literature and we speculate that it may be due to Ref.~\cite{MP99}.
With the help of GW signals associated with compact binaries moving in Newtonian-accurate 
eccentric orbits and inspiralling under quadrupolar RR order, Ref.~\cite{MP99}
demonstrated that the Newtonian order circular templates are quite efficient in extracting 
even their mildly eccentric GW signals.
However, we recently pointed out that the arguments presented in Ref.~\cite{MP99} 
should not be used to support the 
ability of the traditional PN-accurate circular templates
to capture inspiral GWs from compact binaries having tiny orbital eccentricities \cite{TG07}.
This is because in 
Ref.~\cite{TG07}, we demonstrated that 
the three types of circular inspiral templates based on the adiabatic, complete adiabatic
and gauge-dependent completely non-adiabatic approximants, detailed in Ref.~\cite{AIRS} 
and relevant for the circular inspiral under the quadrupolar RR order,
are inefficient in capturing
GWs from eccentric compact binaries having 2.5PN-accurate orbital 
evolution, detailed in Ref.~\cite{DGI}.  
The present paper extends the analysis presented in Ref.~\cite{TG07} by 
incorporating RR effects to 2PN (relative) order, while constructing both the GW signals and 
the circular search templates. 

   In this paper, we explore the ability of a number of PN-accurate circular templates,
detailed in Refs.~\cite{DIS01,AIRS,CC07,AG07}, 
that employ the radiation reaction effects to the relative 2PN accuracy to capture GW signals
from compact binaries in inspiralling eccentric orbits, constructed
by adapting the phasing formalism, detailed in Ref.~\cite{DGI}.
We construct GW signals by perturbing compact binaries 
moving in 3PN-accurate precessing eccentric orbits with 
RR effects that are 2PN-accurate,
beyond the quadrupolar (Newtonian) order.
Further, we employ the orbital binding energy and the time-eccentricity $e_t$,
appearing in the PN-accurate generalized quasi-Keplerian
parametrization for eccentric orbits \cite{MGS},
to characterize the inspiralling and precessing eccentric binaries.
We explore the ability of various 2PN order circular templates to capture 
GW signals from compact binaries in inspiralling eccentric orbits 
by computing the fitting factors (FF), described in Ref.~\cite{DIS}. 
In the present study, the various 2PN order inspiral templates are provided by 
the adiabatic and complete adiabatic Taylor T1, the adiabatic Taylor T4 and the TaylorEt
approximants { [see Eqs.~
(\ref{Eq:P_adT1}),
(\ref{Eq:P_cT1}),
(\ref{Eq:P_T4}) and
(\ref{Eq:P_Et})
 ]}.

  The organization of the paper is as follows. 
In Sec.~\ref{Sec_II:Phasing}, we briefly describe 
the phasing of GWs from 
compact binaries in inspiralling eccentric orbits, available in Ref.~\cite{DGI}.
After that we 
provide certain PN-accurate formulae and relations that 
are necessary to construct 
our GW signals with the help of  
the lengthy PN-accurate expressions, available in Ref.~\cite{KG06}. 
Section~\ref{Sec_II:Phasing} also contains formulae required to construct 
various types of circular templates.
How and why we perform a number of FF computations are explained in Sec.~\ref{Sec_II:PhasingI} along with our 
results. Our conclusions are presented in Sec.~\ref{sec:SIV}.

\section{ Phasing formulae required to model GW signals from eccentric binaries 
 and various circular templates   }
\label{Sec_II:Phasing}

 This section  provides brief descriptions and necessary formulae required to model 
GWs associated with non-spinning compact binaries inspiralling along eccentric 
and circular orbits. For the sake of simplicity, we only consider temporally evolving 
restricted $h_{\times}(t)$, having Newtonian accurate amplitude and PN-accurate phase evolution. 
In the case of compact binaries in inspiralling eccentric orbits, as noted earlier,
we construct $h_{\times}(t)$ by adapting the GW phasing formalism, available in Ref.~\cite{DGI}.
This approach provides an efficient way of modeling 
the PN-accurate orbital dynamics of eccentric binaries  
incorporating the three inherent time scales, 
namely, those associated with
the radial motion (orbital period),
advance of periastron, and
radiation reaction, crucial to construct temporally evolving $h_{\times}(t)$. 
At present, the detailed PN-accurate expressions, describing the orbital dynamics of non-spinning compact
binaries in 3PN-accurate eccentric orbits perturbed by the relative 2PN-accurate 
RR effects, computed with the help of Refs.~\cite{DGI,MGS}, are available in Ref.~\cite{KG06}.
In the next subsection, we summarize what were detailed in Refs.~\cite{DGI,KG06,TG07}
and obtain the restricted $h_{\times}(t)$ associated with non-spinning compact
binaries in precessing and inspiralling eccentric orbits in harmonic gauge. 
Due to the availability of the lengthy PN-accurate expressions in 
Ref.~\cite{KG06}
that are useful to construct 
our fiducial GW signals,we only provide 
certain PN-accurate relations that are crucial 
to implement the the present version of the relevant $h_{\times}(t)$ in the  upcoming subsection. 
However, we do display all the formulae required to construct various types of 
circular templates that are employed in the present study.

\subsection{ GW phasing for compact binaries in inspiralling PN-accurate 
eccentric orbits}

 The dominant quadrupolar contribution to 
the plus polarization, denoted by $h_{\times}\big|_{\rm Q}(t)$,
associated with a compact binary consisting of individual masses $m_1$ and $m_2$ 
at a radial distance $R'$ \cite{DGI} 
reads
\begin{align}
\label{hx_ecc}
h_{\times}(r(t),\phi(t),\dot{r}(t),\dot{\phi}(t)) \big|_{\rm Q}
&=-\frac{2 G m \eta C}{c^4 R'}
\bigg[
\bigg( \frac{G m}{r(t)} + r(t)^2 \dot{\phi}(t)^2 
\nonumber
\\
& \quad
- \dot{r}(t)^2 \bigg) \sin 2 \phi(t)
\nonumber
\\
& \quad
- 2 \dot{r}(t) r(t) \dot{\phi}(t) \cos 2 \phi(t)
\bigg]
\,,
\end{align}
where the symmetric mass ratio $\eta = m_1\,m_2/m^2$ and $m=m_1+m_2$, respectively, 
while $C= \cos i$, $i$ being
the inclination of the orbital plane with respect to the plane of the sky. 
The orbital separation, phase and their time derivatives are denoted by
$r(t),\phi(t),\dot r(t) = d r(t)/dt $ and $\dot \phi(t) = d \phi(t)/dt $, respectively.
 
 The phasing formalism, detailed in Ref.~\cite{DGI}, provides an  
efficient way of implementing the temporal evolution for the dynamical variables,
${r(t), \phi(t), \dot r (t), \dot \phi(t)}$, appearing in Eq.~(\ref{hx_ecc}).
Following Ref.~\cite{DGI}, we note that 
in the case of  compact binaries in inspiralling 
eccentric orbits, however small may be the eccentricity, the time evolution of the dynamical
variables explicitly depends on both the conservative and the reactive contribution 
to the orbital dynamics. 
Due to the existence of a Keplerian type parametric solution to the 3PN-accurate 
conservative orbital dynamics, detailed in Ref.~\cite{MGS}, the conservative contributions to 
$\left \{ r(t), \phi(t), \dot r (t), \dot \phi(t) \right \} $ can be written as
\begin{subequations}
\label{Eq_2:orb_elmts}
\begin{align}
r(t)     &=      r~(u(l), {\cal E}, {\cal J} ) \,,\\
        \dot r(t)     &=      \dot r~(u(l), {\cal E}, {\cal J} )\,, \\
        \phi(t)         &=      \lambda (l; {\cal E}, {\cal J} )  + W(u(l), {\cal E}, {\cal J})\,, \\
        \dot \phi (t)   &=      \dot \phi~(u(l), {\cal E}, {\cal J}) \,,
\end{align}
\end{subequations}
where $u$ and $l$ are the eccentric and mean anomalies associated with the 3PN-accurate 
generalized quasi-Keplerian
parametrization, respectively, while ${\cal E}$ and ${\cal J}$  stand for the orbital energy and 
angular momentum, respectively.  
The angular variable $\phi$ is split into two parts to make sure that there exists a part that 
depends linearly on $l$. 
Recall that such a split allows one to explicate easily the effect of the periastron advance, 
while constructing the frequency spectrum associated 
with $h_{\times}(t)$ relevant for compact binaries in 
PN-accurate eccentric orbits[see Ref.~\cite{TG06} for details].
The angular type variables $\lambda$ and $l$ can be expressed as
\begin{subequations}
\label{Eq_2.L_&_lambda}
\begin{align}
l & \equiv n ( t- t_0) + c_l = l( u;  {\cal E}, {\cal J},c_l )\,,\\
\lambda & \equiv ( 1 + k ) n ( t- t_0) + c_{\lambda} =  \lambda ( t- t_0; {\cal E}, {\cal J}, c_{\lambda})\,,
\end{align}
\end{subequations}
where the constants 
$t_0$, $c_l$ and $ c_{\lambda}$ refer to some initial instant and the values
of $l$ and ${\lambda}$ at $t= t_0$. 
The symbols $n$ and $k$ represent the mean-motion and the measure of advance of the periastron
in the time interval $2\,\pi/n$, respectively.
It turns out that the PN-accurate expressions for $n$ and $k$ in terms of ${\cal E}$ and ${\cal J}$
are the only two gauge-invariant quantities present
in the PN-accurate Keplerian-type parametric solution \cite{DS88,MGS}.
If one ignores the reactive contributions to the orbital dynamics, the explicit temporal evolutions
for the dynamical variables,
$ r(t), \phi(t), \dot r (t), \dot \phi(t) $, are obtained by solving 
the 3PN-accurate Kepler equation (KE) that provides the link between $l$ and $u$.   
The 3PN-accurate KE can be symbolically expressed as
\begin{align}
\label{Eq_2:KE_0123}
l = & u - e_t\, \sin u + l_{\rm 2,3} (u, {\cal E}, {\cal J})\,,
\end{align}
where $l_{\rm 2,3} (u, {\cal E}, {\cal J})$ stands for the 2PN and 3PN corrections to the 
usual Newtonian KE $l= u - e_t\, \sin u$.
With the help of Eq.~(\ref{Eq_2:KE_0123}), we observe that the PN-accurate conservative dynamics,
for example, specified
by the 3PN-accurate  relative acceleration ${\mathcal{A}^{i}}$
in harmonic gauge\cite{LB_lr}, have {\it four} constants of 
integration, namely ${\cal E}, {\cal J}, c_l $ and $c_{\lambda}$.

 To describe the dynamics of compact binaries in inspiralling eccentric orbits, it is essential to include 
the reactive contributions to ${\mathcal{A}^{i}}$ that first enters at the 2.5PN (absolute) order.
However, it is imperative to invoke the energy and angular momentum balance
argument to implement the higher order PN-accurate reactive evolution of an eccentric orbit.
This is because 
the reactive contributions to ${\mathcal{A}^{i}}$ are available only to the 
1PN relative (or 3.5PN absolute) order \cite{LB_lr}. 
Recall that the balance argument equates the orbital averaged
higher order PN-accurate energy and angular momentum fluxes in GWs
to the time derivatives of PN-accurate orbital energy and angular momentum, respectively.
Invoking the  balance argument and 2PN-accurate far-zone energy and angular 
momentum fluxes, available in Refs.~\cite{BS93,RS97,GI97}, it is possible to 
incorporate the 2PN-accurate secular RR effects on  
PN-accurate eccentric orbits in the following way[see Refs.~\cite{DGI,KG06} for details].
To describe the inspiral dynamics of compact binaries, 
Ref.~\cite{DGI} employed an improved `method of variation of constants'    
to include the RR effects onto the conservative orbital dynamics.
The idea is to allow the dynamical variables, appearing in Eq.~(\ref{hx_ecc}), 
have the same functional forms as given by
{ Eqs.~(\ref{Eq_2:orb_elmts}) and (\ref{Eq_2.L_&_lambda})} 
even when the dynamics is not conservative, but reactive.
However, one lets the associated 
constants of integration, namely ${\cal E}, {\cal J}, c_l $ and $c_{\lambda}$ to vary with time.
The equations that govern the temporal evolutions of these quantities are
given by Eqs.~(35) in Ref.~\cite{DGI} and, as expected,
depend on the reactive contributions to ${\mathcal{A}^{i}}$.
Further, the temporal variations of these four constants of
integration can be modeled to consist of a slow drift and fast oscillations, which 
symbolically reads 
\begin{align}
\label{phasing_eq:18}
c_\alpha (l) & =
\bar{c}_\alpha (l) + \tilde{c}_\alpha (l)
\,,
\end{align}
 where $\alpha$ stands for one of the four quantities ${\cal E}, {\cal J}, c_l $ and $c_{\lambda}$. 
In the above equation,
$\bar{c}_\alpha (l)$ denotes the slow drift,
which accumulates over the RR time scale
to induce large changes in $c_\alpha (l)$.
The fast oscillations in $c_\alpha (l)$
are denoted by $\tilde{c}_\alpha (l)$,
and the explicit expressions for 
$d \tilde{c}_\alpha /dt$ are only available to the relative 1PN order \cite{KG06}. 
To the reactive PN order we are interested, it turns out that  
$ d \bar {c_l} /dt = d \bar {c_{\lambda}} /dt  \equiv 0$, while  
$ d \bar {\cal E} /dt  $ and $ d \bar {\cal J} /dt  $ are provided by the 
far-zone energy and angular momentum fluxes, respectively \cite{DGI}.
Further, it was demonstrated in Refs.~\cite{DGI,KG06} that the rapidly oscillating parts have 
substantially smaller amplitudes. Therefore, for the present study,
we ignore the RR induced $\tilde{c}_\alpha (t)$ contributions
to the orbital dynamics.

 The strategy to obtain $h_{\times} \big|_{\rm Q}(t) $ associated with 
compact binaries moving in 3PN-accurate eccentric orbits that are perturbed by
the relative 2PN-accurate RR effects is the following.
With the help of the 3PN-accurate generalized quasi-Keplerian
parametrization in harmonic coordinates, available in Ref.~\cite{MGS}, and Refs.~\cite{DGI,KG06},
we compute the 3PN-accurate expressions for $r, W, \dot r,\dot \phi $ in terms of 
${\cal E}, {\cal J}$ and $u$. This is supplemented by the 3PN-accurate expression for 
$\lambda$ in terms of $l, {\cal E}, {\cal J}$. 
We employed a modified 
version of Mikkola's method \cite{SM87}, detailed in Ref.~\cite{TG07}, 
to solve accurately and efficiently
the 3PN-accurate KE and obtain the temporal evolution for $r(t), W(t), \dot r (t),\dot \phi (t)$  
under the 3PN-accurate conservative orbital dynamics.
Afterwards, we numerically solve the 2PN-accurate differential equations for 
${\cal E}$ and ${\cal J}$ and impose the resulting reactive evolutions in 
${\cal E}$ and ${\cal J}$ onto the earlier computed 3PN-accurate conservative orbital dynamics.
The resulting $r(t), \phi(t), \dot r (t),\dot \phi (t)$  leads to, 
via Eq.~(\ref{hx_ecc}),  $h_{\times} \big|_{\rm Q}(t)$ associated with 
non-spinning compact binaries 
inspiralling along 3PN-accurate eccentric orbits under the action of 
2PN-accurate reactive dynamics.  
  
 Until now, we have employed the orbital energy and angular momentum to describe the eccentric 
inspiral. However, a measure of the non-circularity is better described in terms of 
any of the eccentricity 
parameters, appearing in the PN-accurate Keplerian-type parametric solution \cite{MGS}, 
than in terms of ${\cal J}$.
Therefore, from here onwards, we employ the orbital binding energy ${\cal E}$ and
the time eccentricity 
parameter $e_t$, associated with the PN-accurate KE, to specify our PN-accurate eccentric binaries.   
With the help of Ref.~\cite{KG06}, it is fairly straightforward to obtain 
3PN-accurate expressions for $r, \dot r, \dot \phi, W, \lambda$ and 3PN-accurate KE in terms of 
${\cal E}$ and $e_t$. For the sake of brevity, we provide explicitly the above 
expressions to 1PN order only. However, we do provide the 3PN-accurate relation 
connecting $n$ to ${\cal E}$ required to obtain the lengthy 2PN and 3PN corrections to
$r, \dot r, \dot \phi, W, \lambda$ and $l(u)$ in terms of ${\cal E}$ and $e_t$ from 
Eqs.~(23)-(27) in Ref.~\cite{KG06}.

   It should be noted that the dynamical variables enter $h_{\times} \big|_{\rm Q}(t) $
in certain combinations like $\frac{G\,m}{c^2\, r}, \frac{ r\, \dot \phi}{c} $ and 
$ \frac{\dot r}{c} $. Therefore, we list below the following dimensionless 
dynamical variables required to describe the dynamical evolution of $h_{\times} \big|_{\rm Q}(t)$ 
in a partially symbolic manner. 
\begin{widetext}
\begin{subequations}
\label{Eq.3}
\begin{align}
\label{Eq:sc_Z}
\frac{G\, m}{c^2\,r} (\xi, e_t, u)
 &= \frac{\xi}{\chi} \,
	 \left\{1+	\xi \, \frac{1}{4 \chi}	\left[\chi (9-5 \eta )+6 \eta -16 \right] 
 + \xi^2 (...) + \xi^3 \left (.. \right )
\right\}^{-1}
\\
\label{Eq:dot_r}
\frac{ \dot r }{c} (\xi, e_t, u) &=
 \frac	{\xi^{1/2} \, e_t \, \sin u}
		{\chi } \,
		\left \{   1 + \left[ \frac{3}{8} -\frac{9}{8} \eta  \right] \xi  
 + \xi^2 (...) + \xi^3 \left (.. \right )
  \right \}
\\
\label{Eq:r_omg_c}
\frac{\, r (\xi, e_t, u)  \times { \dot \phi } (\xi, e_t, u)}{c}
 &=	\frac{\xi^{1/2} \, \sqrt{1-e_t^2}}{\chi}
	 \times
	\left \{  1+ \xi \, \frac{1}{4 \chi}	\left[ \chi (9-5 \eta )+6 \eta -16 \right] 
 + \xi^2 (...) + \xi^3 \left (.. \right )
\right\}
	\times
 \biggl \{
	1
\nn \\ &
+\xi \frac{1}{8 (1-e_t^2)}
		 \big [ 
					-23 + 9 \eta + e_t^2 (15 - \eta) 
		+	\frac{1}{\chi}	\biggr ( 32 - 8 \eta + e_t^2 (-32 + 8 \eta) \biggr )
 + \xi^2 (...) + \xi^3 \left (.. \right )
		 \big ]
 \biggr\}
\\
\label{Eq:phi}
\phi ( \xi, e_t, u, l)
&= \lambda (l, \xi, e_t, ) + W(\xi, e_t, u) \,, \mbox{where}
\\
\label{Eq:lambda}
\lambda (l, \xi, e_t, )
&=
	\xi^{3/2}
	\frac{c^3}{G \, m}
	\times
	\left\{ 1 + \xi \left[\frac{1}{8}(-15 + \eta)
			+\frac{3}{(1-e_t^2)} \right]
+ \xi^2 (...) + \xi^3 \left (.. \right )
 \right\} \,(t-t_0)
\\
\label{Eq:W}
W(\xi, e_t, u)|_{1PN} ~=&~
	(v-u) ~+~ e_t \sin u ~+~
	\xi \,
	\frac{3}{(1-e_t^2)} \,
	\left\{	(v-u) + e_t \, \sin u \right\} 
+ \xi^2 (...) + \xi^3 \left (.. \right )
\end{align}
\end{subequations}
\end{widetext}
where $\xi = -2\, {\cal E}/\mu\,c^2 $, $\mu$ being the reduced mass of the binary: $\mu = \eta\, m$ 
and $\chi = (1 -e_t\, \cos u )$.
To evaluate $W(\xi, e_t, u)$, we employ the following expression for $v-u$
\begin{align}
\label{phasing_eq:app_vmu_eq1}
v - u & = 2 \tan^{-1}
\left(
\frac{ \beta_{\phi} \sin u }{ 1 - \beta_{\phi} \cos u }
\right)
\,,
\end{align}
where $\beta_{\phi} = ( 1 - \sqrt{ 1 - e_\phi^2 } ) / e_\phi$.
Using the PN accurate relation connecting $e_{\phi}$ to $e_t$, extractable from Ref.~\cite{MGS},
it is fairly straightforward to express $\beta_{\phi}$ in terms of $e_t$ and $\xi$.  To 1PN order,
$\beta_{\phi}$ explicitely reads
\begin{widetext}
\begin{align}
\label{phasing_eq:app_vmu_betaph}
\beta_{\phi}(\xi, e_t, \eta) ~=&~
	\frac{1-\sqrt{1-e_t^2}}{e_t}
	\biggl \{  1 + \xi \frac{(4 -\eta) }{\sqrt{1-e_t^2}} 
+ \xi^2 (...) + \xi^3 \left (.. \right )
\biggr \} 
\end{align}
\end{widetext}

   The explicit expressions for the 2PN and 3PN contributions to $ G\, m/c^2\,r, \dot r/c, \lambda,  W, 
r\, \dot \phi/c $ and $\beta_{\phi}$,
symbolically noted as `$+ \xi^2 (...) + \xi^3 \left (.. \right )$' in their respective 1PN-accurate expressions, 
are obtainable in a straightforward manner using Eqs.~(23)-(26) in Ref.~\cite{KG06} 
and the following 3PN-accurate expression connecting $n$ to $\xi$ and $e_t$\cite{MGS}
\begin{widetext}
\begin{align}
\label{Eq:n_xi}
n 	&=	\xi^{3/2} \frac{c^3}{G \, m}
			\Biggl \{
			1	- \xi	\frac{1}{8} \biggl[15 - \eta \biggr]
 				+ \xi^2 \frac{1}{128} \,
					\biggl [
					( 555 + 30 \eta + 11 \eta ^2 )  
- \frac{1}{\sqrt{1-e_t^2}} \,  (960 - 384 \eta )
					\biggr ]
\nn \\&
+ \xi^3 \,\frac{1}{3072 (1-e_t^2)^{3/2}}
                        \Biggl [
                16 \,   \left (-2340 + \{10244 - 123 \pi^2\} \eta - 792 \eta^2 -
                        36 \, e_t^2 \{255 - 159 \eta + 34 \eta^2\}
                        \right )
                + \nn \\ &
                (1-e_t^2)^{3/2} \,
                        \left (
                        45 \, (-653 - 111 \eta - 7 \eta^2 + 3 \eta^3)
                        \right )
                        \Biggr ]
			 \Bigg \}
\end{align}
\end{widetext}
 
  The explicit conservative temporal evolution for the above listed orbital variables are obtained 
by solving the 3PN-accurate KE connecting $u$ to $l$. The right hand side of the PN-accurate KE reads 
\begin{widetext}{
\bea
\label{Eq:KE3PN}
l(\xi, u, e_t)	=&&
		u-e_t \sin u + 
		\xi^2	\biggl \{
		\frac{\left(\frac{15}{2}-3 \eta \right) (v-u)}{\sqrt{1-e_t^2}}
+\frac{e_t (15-\eta ) \eta  \sin (u)}{8 \chi}
			\biggr\} 
+ \xi^3 \left (.. \right )
\eea
}
\end{widetext}
where the $v-u$ term appearing at the 2PN order on the right hand side of Eq.~(\ref{Eq:KE3PN}) 
is evaluated using 1PN-accurate expression for $\beta_{\phi}$.
As noted earlier, for an accurate and efficient 
determination of $u(l)$, we employed a modified version of Mikkola's method for solving
the classical KE \cite{SM87}(see Ref.~\cite{TG07} for its implementation at the 2PN order).
With the help of PN-accurate $u(l)$, it is fairly easy to obtain 
$h_{\times} \big|_{\rm Q}(t) $ associated with compact binaries evolving under 3PN-accurate 
conservative orbital dynamics.

 Let us now impose the effects of 2PN-accurate RR onto the above detailed conservative 
orbital dynamics. Due to the neglect of the small amplitude and 
fast oscillatory contributions to ${\cal E} $,  
$ {\cal J} $, $c_l$ and $c_{\lambda}$,  the reactive orbital evolution is entirely given by 2PN-accurate 
orbital averaged expressions for the energy and angular momentum fluxes, and these 
PN-accurate expressions in harmonic gauge can be extracted from Refs.~\cite{BS93,RS97,GI97}. 
However, in this study, the PN-accurate eccentric orbits are specified 
in terms of $\xi$ and $e_t$, and hence require the 2PN-accurate differential
equations for $\xi$ and $e_t$, written in terms of  $\xi$ and $e_t$.
Due to the definition of $\xi$,
the differential equation governing the reactive 
evolution of $ \xi$ follows directly from the 2PN-accurate far-zone energy flux. 
The differential equation describing the secular 2PN-accurate reactive evolution of $ e_t$,
requires i) the 2PN accurate expression for $e_t^2$ in terms of ${\cal E}$ and ${\cal J}$ in harmonic gauge, 
available in Ref.~\cite{MGS}, ii) the 2PN-accurate expressions for the 
orbital averaged far-zone energy
and angular momentum fluxes and finally iii) the energy and angular momentum balance argument.
The resulting 2PN-accurate expressions for $d \xi/dt $ and $d  e_t/dt$, describing the secular 
evolutions of $\xi$ and $e_t$, is given by 
\begin{widetext}{
\begin{subequations}
\begin{align}
\label{Eq:dXidt}
\frac{d  \xi}{d t}
      &= 2\, \frac{\xi^5 \eta ^2}{15} \, \frac{c^3}{G\,m}
	\left \{	\dot { \xi}^{N}
		+	\dot { \xi}^{1PN}
		+	\dot { \xi}^{1.5PN}
		+	\dot { \xi}^{2PN}
	\right\}\,, \\
\frac{d  e_t}{d t}
      &= - \frac{\xi^4\,e_t\, \eta }{15} \, \frac{c^3}{G\,m}
	\left \{	\dot { e_t}^{N}
		+	\dot { e_t}^{1PN}
		+	\dot { e_t}^{1.5PN}
		+	\dot { e_t}^{2PN}
	\right\}\,, \mbox {where} \\
\dot { \xi}^{N}
	&=	\frac{1}{\left(1-e_t^2\right)^{7/2}}	\left\{96 + 292 e_t^2 + 37 e_t^4\right\} \\
\dot { \xi}^{1PN}
	&=	\frac	{\xi}
			{56 \left(1-e_t^2\right)^{9/2}}
			{\big \{
							   208-13440 \eta
					+	8	[22340-19607 \eta ]  e_t^2
					+	42	[ 5966-3459 \eta  ]  e_t^4
					+		[19487-8806 \eta  ]  e_t^6
		\big \}		}								\\
\dot { \xi}^{1.5PN}
	&=	\frac{\xi^{3/2} \pi}{96}
			 \left\{	  36864
				+ 448320 e_t^2
				+ 2061840 e_t^4
				+ 6204647 e_t^6
			\right\}\,,								\\
\dot { \xi}^{2PN}
	&=
\frac{\xi^2}{6048 (\sqrt{1-e_t^2})^{11}}
		\big \{
		32 \left[253937	 - 162585   \eta + 45360   \eta ^2 \right]
		+16 \left[2007326	 - 6275466  \eta + 3509541 \eta ^2 \right] e_t^2 \nn \\
	&~	+12 \left[22179502	 - 35103402 \eta + 13897611\eta ^2 \right] e_t^4
		+18 \left[12638659	 - 12297567 \eta + 3878616 \eta ^2 \right] e_t^6 \nn \\
	&~	+9  \left[1374197	 - 916728   \eta + 254856  \eta ^2 \right] e_t^8
		-9072 \left[
					96
				-	1060	e_t^2
				-	1863	e_t^4
				-	148	e_t^6
			\right] \sqrt{1-e_t^2} (5 - 2 \eta)
		\big \}\,,
 \\
\dot { e_t}^{N} &= \frac{\left(121 e_t^2 + 304\right) }{ (\sqrt{1-e_t^2})^5}
\,,\\
\dot { e_t}^{1PN} &=  \frac   { \xi} {56\, (\sqrt{1-e_t^2})^7} 
\biggl [ 
	\left\{75667 e_t^4+344784 e_t^2-980 \left(34 e_t^4+225 e_t^2+72\right) \eta -28536 \right\}
\biggr ]\,,\\
\dot { e_t}^{1.5PN} &=  \frac   { \xi^{3/2}\, \pi } {96 } 
\biggl [
189120+1042992\,{{e_t}}^{2}+3061465\,{{ e_t}}^{4}
\biggr ]\,, \\
\dot { e_t}^{2PN} &=  \frac   { \xi^{2}\, } {2016\, (\sqrt{1-e_t^2})^9 } 
\biggl [
\left( 17914878-12062097\,\eta+3288852\,{\eta}^{2} \right) {{ e_t }}^{6}
\nn \\ &
+ \left( 128983907  -139206105\,\eta  +45633420\,{\eta}^{2}  \right) 
{{ e_t}}^{4}
+ \left( 46339524\,{\eta}^{2}-94111416\,\eta+28071312
 \right) {{ e_t}}^{2}
\nn \\ &
+7116856
-6772968\,\eta +3566304\,{\eta}^{2}
+3024\, \left( 511\,{{ e_t}}^{4}+2809\,{{e_t}}^{2}+80
 \right)  \left( 5-2\,\eta \right) \, \sqrt{1-e_t^2}
\biggr ]\,.
\label{Eq:detdt}
\end{align}
\end{subequations}
}
\end{widetext}
The contributions entering at the relative 1.5PN order are due to the
dominant order GW tails that 
are usually expressed in terms of certain infinite sums 
involving the Bessel functions $J_{p}(p\,e_t)$ and its derivative 
$J'_{p}(p\,e_t)$
\cite{BS93,RS97}.
In the above equations, we have expanded those infinite sums to ${\cal O}(e_t^4)$. 
This is justified as we are only interested in eccentric binaries having initial $e_t \leq 0.2$
in the present study. 
Let us remind the reader that it is due to the neglect of the {\it tilde} contributions to
${\cal E}, e_t, c_l$ and $c_{\lambda}$ that the reactive evolution is fully 
prescribed by the above differential equations for $ \xi (t) $ and $ e_t$.
And, the above two equations represent the secular variations in 
$\xi$ and $e_t$, namely $\bar \xi (t)$ and $\bar e_t (t)$, due to the emission of GWs.

 We are now in a position to implement numerically
$h_{\times} \big|_{\rm Q}(t)$ emanating from 
non-spinning compact binaries
inspiralling along 3PN-accurate eccentric orbits under the action of
2PN-accurate reactive dynamics. 
 It is obvious that we need to specify 
the initial and final values for $\xi$ so that 
the dominant GW spectral component of $h_{\times} \big|_{\rm Q}(t)$ 
evolves in the frequency window $40$Hz - $\left( 6^{3/2}\, \pi\,  G m/c^3 \right )^{-1 } $ Hz, relevant for the
initial LIGO. 
We compute the bounding values for $\xi$ with the help of Ref.~\cite{TG06}  that 
demonstrated the fact that  
the dominant GW spectral component
of a mildly eccentric compact binary, having PN accurate orbital motion, appears at
$\frac{1}{\pi}\, <\frac{d \phi}{dt}> \equiv \frac{n\,(1 + k)}{\pi}$, where 
$<\frac{d \phi}{dt}>$ stands for the PN-accurate orbital average of 
$d \phi/dt$.
The PN-accurate expression for $<\frac{d \phi}{dt}>$
can be computed using the procedure, detailed in
Ref.~\cite{BS89}, involving the Laplace second integrals for 
the Legendre polynomials[see Eqs.~(4.16)-(4.19) in Ref.~\cite{BS89}] .
For the present paper, we use the following 3PN-accurate expression for 
the orbital averaged orbital angular velocity
$<\frac{d \phi}{dt}>$ 
to obtain the bounding values for $\xi$ for a compact binary specified by
certain $m,\eta$ and $e_t$ values:
\begin{widetext}
{
\bea
\label{Eq:av_dot_phi}
\left< \frac{d \phi}{dt} \right>  &\equiv&
{n (1+k)} 
\nonumber \\
	&=&	\xi^{\frac{3}{2}} 
			\frac	{c^3}
				{G \, m}
\,		\Biggl\{
		1
		 +		 \xi
				\left[  \frac{\eta }{8}+\frac{3}{1-e_t^2}-\frac{15}{8} \right]
		 +		\frac{\xi^2}{128 \, ( 1-e_t^2)^2}
				\Biggl[
				              (1851 - 786 \, \eta + 11 \, \eta^2)
				 - 2 \, e_t^2 (-861 + 486 \, \eta + 11 \, \eta^2) \nn \\
		&&		 +      e_t^4 ( 555 +  30 \, \eta + 11 \, \eta^2)
				+ 192 (1 - e_t^2)^{3/2} \, (-5 + 2 \eta)
				\Biggr]
		+		\frac{\xi^3}{3072 \, (1-e_t^2)^3}
				\Biggl[
					 3 \, (76965 + (-167089 + 3936 \pi^2) \, \eta  \nn \\ && + 5439 \, \eta^2 + 45 \, \eta^3)
				+	 3 \, e_t^2 (228249 + (-222781 + 984 \pi^2) \, \eta + 40491 \, \eta^2 - 135 \, \eta^3)
				+	 9 \, e_t^4 ( 25301 - 17201 \eta + 3471 \eta^2  \nn \\ && + 45 \eta^3)
				-	45 \, e_t^6 (  -653 -   111 \eta - 7 \eta^2 + 3 \eta^3)
				+	\sqrt{1-e_t^2}
					\big(
					16       (-6660 - 41 (-292 + 3 \pi^2) \eta - 792 \eta^2  \nn \\
		&&		+	   e_t^2 ( 6120 + (-9704 + 123 \pi^2) \eta - 432 \eta^2)
				+	36 e_t^4 (  255 - 159 \eta + 34 \eta^2))
					\big)
		\Biggr]
\Biggr\}
\eea
}
\end{widetext}
With these inputs, it is fairly straightforward to construct $h_{\times} \big|_{\rm Q}(t)$
for inspiralling compact binaries, having some tiny orbital eccentricity, 
emitting GWs in the initial-LIGO frequency window.

 Finally, let us explain why we employed $\xi $, along with $e_t$, to represent 
the PN-accurate eccentric orbit, which guided us to use $\xi$ as our PN expansion parameter. 
In the literature, the GW phasing for eccentric binaries are 
performed using either $\left (\frac{G\,m\,n}{c^3} \right)^{(2/3)}$ 
\cite{DGI,KG06} or $\left (\frac{G\,m\, \omega}{c^3} \right)^{(2/3)}$ \cite{ABIQ,Hinder08}, where 
$\omega $ stands for the above listed  3PN-accurate $<\frac{d \phi}{dt}>$, 
as the PN expansion parameters, along with $e_t$. 
The choice of $n= 2\,\pi/T_r$, where the radial period $T_r$ being the time of return to the periastron,
as a PN expansion parameter may be visually 
problematic. This is because it is rather difficult to
pinpoint the periastron position for circular binaries, making it difficult to  
imagine the meaning of $T_r$ in the case of circular inspiral.      
We speculate that this may be the one of the reasons for 
Ref.~\cite{ABIQ} to introduce $\omega \equiv <\frac{d \phi}{dt}> = n \, \left ( 1 + k \right )$
as their PN expansion parameter.
However, it is not clear why one should employ the orbital averaged angular velocity to perform GW 
phasing for eccentric binaries. The reason stated in 
 Ref.~\cite{ABIQ} 
for introducing $<\frac{d \phi}{dt}>$ 
as the PN-expansion parameter  is that it generalizes the 
$\omega = d \phi/dt $ parameter used in the construction of the usual PN-accurate inspiral templates,
available in Ref.~\cite{DIS01}.
In other words, the PN accurate expression for $d \phi/dt$, given in terms of 
$\xi,e_t, u$, when re-expressed in terms   
$\omega \equiv <\frac{d \phi}{dt}>, e_t, u$ using Eq.~(\ref{Eq:av_dot_phi})
will lead to  $d \phi/dt = \omega $
in the circular limit [check Eqs.~(A.10)-(A.14) in Ref.~\cite{Hinder08} for the actual demonstration of the 
above statement].  
It is important to realize that 
the use of $d \phi/dt = \omega $ as the PN expansion 
parameter for the circular inspiral is usually justified by the argument that 
it defines the instantaneous orbital angular frequency. 
In the case of the heavily employed 
restricted PN-accurate inspiral templates \cite{DIS01},  the instantaneous GW frequency $f_{\rm GW}$ is 
related to the instantaneous orbital angular frequency by 
 $ f_{\rm GW} \equiv \omega/\pi$.
However, in the case of GW phasing for eccentric binaries,
neither $n$ nor  $\omega \equiv <\frac{d \phi}{dt}> $ 
are sufficient to define the instantaneous orbital angular frequency.
Further, the associated GW spectrum can not be expressed solely in terms of $n$ or
$<\frac{d \phi}{dt}>$ \cite{TG07}. These considerations prompted us to employ 
neither $n$ nor $ \omega \equiv <\frac{d \phi}{dt}>$ 
as our PN expansion parameter for performing the GW phasing for eccentric binaries.   
We would like to stress that, while considering the 
PN-accurate conservative orbital dynamics,  our PN expansion parameter requires no orbital averaging 
or any prior knowledge about the turning points of the associated eccentric orbit. Further,  
$\xi$  provides a measure of the instantaneous orbital energy of an eccentric orbit.
In the case of circular inspiral,
it is also straightforward to identify  $\xi$ as a measure for the instantaneous orbital energy.
Interestingly, for Newtonian circular orbits, our PN expansion parameter
$\xi$ is indeed a measure of $\frac{v^2}{c^2} $ like
$ (\frac{G\,m\,\omega }{c^3})^{2/3}$ and $(\frac{G\,m\,n}{c^3})^{2/3}$ variables available in the literature.

 It was argued in Ref.~\cite{DGI} that the GW phasing for eccentric binaries that employ 
$n$ as the PN expansion parameter is not going to be accurate near the last stable orbit (LSO).
This is one of the reasons for terminating the eccentric orbital evolution around 
$j^2 \sim 48$ in Refs.~\cite{DGI,KG06}, where $j = c\, {\cal J}/G\,m\,\mu$.
Recall that  for a test particle  
around a Schwarzschild black hole, the last stable circular orbit is also given by 
$j^2 = 12$ \cite{DS88}.
A recent comparison between PN and Numerical Relativity (NR) 
 based GW phase evolutions for equal-mass 
eccentric binaries, reported in Ref.~\cite{Hinder08}, also supported the 
above argument. However, Ref.~\cite{Hinder08} demonstrated that the PN-accurate GW 
phasing for eccentric binaries, having  3PN-accurate conservative 
and 2PN accurate reactive dynamics, that employ $\omega \equiv <\frac{d \phi}{dt}> $
as the PN expansion parameter are fairly accurate even near the LSO.
This observation can be linked to the fact that the GW phasing for eccentric binaries in terms of 
$\omega \equiv <\frac{d \phi}{dt}> $ and $e_t$, 
in the limit $e_t \rightarrow 0$ provides the 2PN-accurate TaylorT4 approximant. And, Refs.~\cite{CC07,GHHB} 
recently demonstrated that the GW phase evolution under this TaylorT4 approximant is fairly 
close to its NR counterpart for equal mass binary black holes having around 
10-15 orbital cycles close to their LSOs.
We would like to point out that in the limit $e_t \rightarrow 0$, 
our present prescription  for doing GW phasing for eccentric binaries
leads to the TaylorEt approximant that employ the 3PN accurate expression for 
$d \phi/dt$ and the 2PN accurate expression for $d \xi/dt$ \cite{AG07}.
With the help of  Fig.~4 in Ref.~\cite{GHHB}, we can state that the 
the accumulated GW phase difference 
between the above approximant and NR,
during the late inspiral stage of an equal-mass binary, 
is comparable to what is provided the TaylorT4 approximant at 2.5PN order. 
The above phase difference 
turned out to be substantially smaller than
a similar accumulated GW phase difference originating from a 2PN approximant 
that employs $(\frac{G\,m\,n}{c^3})^{2/3}$ as
the PN expansion parameter \cite{AG07}.  
Therefore, it is reasonable to expect that during the 
late inspiral stage,
our present prescription for doing eccentric 
GW phasing could be closer to NR than what is presented in Refs.~\cite{DGI,KG06}.
We are initiating an effort to make such a comparison 
with the the help of yet-to-be available fully 3PN-accurate RR effects for 
the eccentric inspiral, similar to what is reported in Ref.~\cite{GHHB}.

 In what follows, we briefly describe how we construct various PN-accurate circular inspiral templates
that incorporate RR effects
to the relative 2PN accuracy.

\subsection{Various PN-accurate circular inspiral templates that incorporate RR effects 
at the relative 2PN order}

In this subsection, we briefly describe how to construct the earlier mentioned PN-accurate 
templates relevant for the circular inspiral.
The crucial PN-accurate  quantities that are required to construct these circular inspiral
templates are the 3PN-accurate orbital binding energy ${\cal E} (x)$ \cite{DJS01,LB_lr} and 
the 2PN-accurate GW luminosity
$ {\cal L }(x) $ \cite{BDIWW}, both of which are
usually expressed as PN series in the gauge invariant quantity $x\equiv (G\,m\, \omega/c^3)^{2/3}$,
where $\omega$ is the instantaneous orbital angular frequency of the associated circular binary.
The explicit expressions for ${\cal E }(x)$ and $ {\cal L }(x) $ relevant for the present 
study are

\begin{widetext}
\begin{subequations}
\label{Eq:Phasing_x}
\begin{align}
\label{Eq:Phas_x_Energy}
{\cal E}(x) =~& -\frac{\eta\, m\, c^2}{2}\,x
\biggl \{
1
- \frac{1}{12} \biggl [ 9 +  \eta \biggr ] x
-
\biggl [
{\frac {27}{8}}
-{ \frac {19}{8}}\,\eta
+\frac{1}{24}\,{\eta}^{2}
\biggr ]{x}^{2}
- \biggl [
{\frac {675}{64}}
+ \left( {\frac {205}{96}}\,{\pi}^{2}-{\frac {34445}{576}} \right) \eta
\nn \\ &
+{\frac {155}{96}}\,{\eta}^{2}
+{\frac {35}{5184}}\,{\eta}^{3}
\biggr ]
{x}^{3}
\biggr \}\,,\\
\label{Eq:Phas_x_Flux}
{\cal L}(x) =~& \frac{32\,\eta^2\,c^5}{5\,G}\, x^{5}\,
\biggl \{
1
- \biggl [ {\frac {1247}{336}}+{\frac {35}{12}}\,\eta \biggr ] x
+4\,\pi \,{x}^{3/2}
- \biggl [ {\frac {44711}{9072}}
-{\frac { 9271}{504}}\,\eta
-{\frac {65}{18}}\,{\eta}^{2} \biggr ] {x}^{2} \biggr\}
\,,
\end{align}
\end{subequations}
\end{widetext}

 The time-domain $x$-based circular templates
are also constructed under the 
restricted PN-accurate prescription  and this leads to 
\begin{align}
h_{\times} = -\frac{ 4\, G\, m\, \eta\,C}{c^2\, R'}\,  x (t)\, \sin 2\, \phi (t)\,,
\end{align}
Following Refs.~\cite{DIS01,AIRS}, one can construct different template families 
that provide 
different prescriptions for the evolution of $x(t)$ and $ \phi(t)$ appearing in the above equation.
The 2PN order adiabatic TaylorT1 approximant of Ref.~\cite{DIS01} is given by
\begin{subequations}
\label{Eq:P_adT1}
\begin{align}
\label{Eq:Phas_x_dphdt_1}
\frac{d \phi (t)}{dt} & = \omega (t)  \equiv \frac{c^3}{G\,m}\, x^{3/2}\,,\\
\label{Eq:Phas_x_dxdt_2PN_T1}
\frac{d\,x(t)}{dt} &=  -\frac{{\cal L}( x)}{ \left( d {\cal E}_{\rm 2PN} / d x
					     \right)}\,,
\end{align}
\end{subequations}
where ${\cal E}_{\rm 2PN}$ stands for the 2PN-accurate orbital binding energy, 
obtainable by dropping $x^3$ terms in Eq.~(\ref{Eq:Phas_x_Energy})
 for ${\cal E}(x)$.
It should be noted that in this approximant, available in LAL, one truncates
the PN-accurate expressions for ${\cal E}(x)$ and ${\cal L}(x)$ at the same relative PN order. 
We can also create the complete adiabatic TaylorT1 approximant 
in the following way \cite{AIRS}
\begin{subequations}
\label{Eq:P_cT1}
\begin{align}
\label{Eq:Phas_x_dphdt_2}
\frac{d \phi (t)}{dt} & = \omega (t)  \equiv \frac{c^3}{G\,m}\, x^{3/2}\,,\\
\label{Eq:Phas_x_dxdt_compl_T1}
\frac{d\,x(t)}{dt} &=  -\frac{{\cal L}( x)}{ \left( d {\cal E} / d x
\right)}\,,
\end{align}
\end{subequations}
where ${\cal E}(x)$ is indeed 3PN-accurate and is given by Eq.~(\ref{Eq:Phas_x_Energy}).

 The TaylorT4 approximant, introduced in Ref.~\cite{CC07}, is obtained by Taylor expanding 
in $x$ the right-hand-side of either
Eq.~(\ref{Eq:Phas_x_dxdt_2PN_T1}) or Eq.~(\ref{Eq:Phas_x_dxdt_compl_T1}) to the 2PN order in $x$.
The 2PN-accurate TaylorT4 is defined by

\begin{widetext}
\begin{subequations}
\label{Eq:P_T4}
\begin{align}
\label{Eq:Phas_x_dphdt}
\frac{d \phi (t)}{dt} =~& \omega (t)  \equiv \frac{c^3}{G\,m}\, x^{3/2}\,,\\
\label{Eq:Phas_x_dxdt_2PN_T4}
\frac{d\,x(t)}{dt} =~& 
\frac{c^3}{G\,m}\, \frac{64\,\eta}{5}\, x^5
\biggl \{
1
- \left( {\frac {743}{336}}+ \frac{11}{4}\,\eta \right) x
+4\,\pi\,{x}^{3/2}
+ \left( {\frac {34103}{18144}}+{\frac {13661}{2016}}\,\eta+{\frac {59}
{18}}\,{\eta}^{2} \right) {x}^{2}
\biggr \}
\,.
\end{align}
\end{subequations}
\end{widetext}

 While searching for inspiral GWs, the LSC employs (or has employed even in the recent past)
the Fourier domain 2PN-accurate
TaylorF2 approximants \cite{lsc_grb}. This approximant can be constructed 
analytically 
from the above TaylorT4 approximant
using the stationary phase approximation \cite{BO} making it computationally cheaper
than the time-domain TaylorT4 approximant.  However, the focus of the present study is not   
the computational costs associated with the implementation of various template families, 
therefore, we employed the 2PN-accurate time-domain TaylorT4 approximant rather than 
its Fourier domain counterpart. 
We observe that among the three approximants,
the LSC Algorithm Library (LAL)\cite{LAL}     
strictly provides only one of them, namely, the adiabatic TaylorT1 approximant at the 2PN order.

  Finally, we construct the 2PN version of the TaylorEt using Ref.~\cite{AG07} as

\begin{widetext}
\begin{subequations}
\label{Eq:P_Et}
\begin{align}
\label{Eq:Phas_Et_hx}
h(t) &=  -\frac{ 4\, G\, m\, \eta\,C}{c^2\, R'}\, \xi (t)\, \sin 2\,\phi(t)
\,,\\
\label{Eq:Phas_Et_dphidt}
\frac{d \phi (t)}{dt} \equiv~& \omega (t) = \frac{c^3}{G\,m}\, \xi^{3/2}
\biggl \{
1
+ {\frac {1}{8}}
\left( {9}+\eta \right) \xi
+ \biggl [
{\frac {891}{128}}
-{\frac {201}{ 64}}\,\eta
+{\frac {11}{128}}\,{\eta}^{2}
\biggr ]
{\xi}^{2}
\no
&+ \biggl [
{\frac {41445}{1024}}
- \biggl ( {\frac {309715}{3072}}
-{\frac {205}
{64}}\,{\pi}^{2} \biggr) \eta
+{\frac {1215}{1024}}\,{\eta}^{2}
+{\frac {45}{1024}}\,{\eta}^{3}
\biggr ]
 {\xi}^{3}
\biggr \}
\,,\\
\label{Phas_Et_dXidt}
\frac{d\,\xi (t)}{dt} =~&
\frac{c^3}{G\,m}\,
{\frac {64}{5}}\,\eta\,{\xi}^{5}
\biggl \{
1
+ \left( {\frac {13}{336}}- \frac{5}{2}\,\eta \right) \xi
+4\,\pi\,{\xi}^{3/2}
+ \left(
{\frac {117857}{18144}}
-{\frac {12017}{2016}}\,\eta
+\frac{5}{2}
\,{\eta}^{2} \right) {\xi}^{2}
\biggr \}
\,,
\end{align}
\end{subequations}
\end{widetext}

 The above approximant obviously provides the circular limit of GW phasing for eccentric inspiral, detailed in the 
previous subsection, and hence employ $\xi$ as the PN expansion parameter. 
A number of features of this approximant are worth mentioning.  
The TaylorEt approximant is specified by
two separate PN-accurate differential equations and in contrast,
the various $x$-based approximants are governed by
just one PN-accurate differential equation. 
In a given GW frequency band, 
the limits of the integration for
various $x$-based approximants,
irrespective of whether one is
interested in the Newtonian or 3.5PN order approximant,
are the same and are independent of the symmetric mass ratio $\eta$.
However, the limits of the integration for the TaylorEt approximant
do depend on the underlying 3PN-accurate conservative dynamics and
therefore depend on $\eta$.

 The plots of $H_{\times}(t)$ representing $h_{\times}\big|_{\rm Q}(t)$ scaled by $ G\,m\, \eta \,C/c^2\,R'$
evolving under various prescriptions detailed in this section are displayed in 
Fig.~\ref{FIG:h_c_e_T1_T4_Et}.
These plots, displaying the evolving $H_{\times}(t)$  in the initial-LIGO frequency window, 
represent the canonical neutron star- black hole binary. 
The plots are for 
i) an eccentric binary inspiralling 
with an initial $e_t = 10^{-3}$, ii) the TaylorEt approximant, iii) the adiabatic TaylorT1
approximant and iv) the TaylorT4 approximant. The similar duration for the eccentric
and the circular TaylorEt $H_{\times}(t)$ should be noted.

 Let us now move on to explore the ability of various circular templates to capture GWs
from inspiralling compact binaries having tiny orbital eccentricities.

\section{Performances of various circular templates against GW signals from compact binaries in 
inspiralling eccentric orbits }
\label{Sec_II:PhasingI}

 In this section, we explore how effectual and faithful 
are the various 2PN order circular templates,
introduced in the previous section, with respect to GW signals from compact binaries
in inspiralling eccentric orbits that should be interesting to the initial-LIGO data analysts.
This is achieved by computing the fully maximized overlaps involving 
the eccentric $h_{\times}(t)$, treated to be the fiducial inspiral GW signal, and 
the \emph {four} types of circular inspiral templates, given by
Eqs.~(\ref{Eq:P_adT1}),
(\ref{Eq:P_cT1}),
(\ref{Eq:P_T4})
and (\ref{Eq:P_Et}).
We extensively employ Ref.~\cite{DIS} to compute the fully maximized overlaps with the help of  
their minimax overlap.
In our numerical experiments, the maximization over the time-lags are straightforward
as we invoke the fast-Fourier-transform routine of the Numerical Recipes \cite{NumReci}
to compute the Fourier-domain versions of our time-domain GW signals and templates.
Further, the maximization over $m$ and $\eta$, characterizing the various circular templates are 
performed using the {\it amoeba} routine\cite{NumReci}.
In the present context, a given circular template family is considered to be 
effectual and faithful if some members of the family can provide  
fully maximized overlaps, hereon the fitting factors (FFs), that are
closer to unity and have smaller biases
in estimating the parameters representing our fiducial GW signal.
In other words, the values of $m$ and $\eta$ characterizing the circular template having the largest FF
provide a measure of the faithfulness of that particular template family.
It turns out that the drop in event rate is proportional to 
$(1 - {\rm FF}^3)$ \cite{FF_er} and therefore, it is 
indeed desirable to have circular template families having 
FF values that are  $\ge 0.97$ with respect to 
fiducial GW signals from compact binaries having tiny
orbital eccentricities. This is mainly because the expected astrophysical GW sources
will have tiny, but non-zero, orbital eccentricities.

  In Table~\ref{Table:XC_T4(2+2)}, we show the FFs involving our fiducial eccentric GW signals
and the 2PN order TaylorT4 circular templates, 
 relevant for the initial LIGO.
We restrict our attention to the usual canonical binaries representing 
neutron star--neutron star, black hole--neutron star and
black hole--black hole binaries. In our FF computations,  
we evolve both our fiducial GW signals and circular templates in the frequency range 
$40$ Hz to $ \left( 6^{3/2}\,\pi\, G\,  m /c^3\right )^{-1}$Hz. 
The numbers clearly indicate that the 2PN TaylorT4 circular templates are 
neither effectual nor faithful in extracting our fiducial GW signals from inspiralling 
binaries, having low orbital eccentricities like $e_t \sim 0.001$.
The numbers associated with the $e_t = 10^{-3}$ case 
are representative of what 
to expect for still lower values for the initial $e_t$.
We justify the above statement by repeating the FF computation employing
the circular TaylorEt approximant to model 
the fiducial GW signal.  
And, the resulting
numbers are listed in the first row of Table \ref{Table:XC_T4(2+2)}.

 Next in Table \ref{Table:XC_AdT1_CAdT1}, we show the results of our FF computation
involving the adiabatic and complete adiabatic TaylorT1 circular
templates that require 2PN accurate RR effects.
The complete adiabatic TaylorT1 templates are invoked so that we can include 
the effects of 3PN-accurate conservative orbital dynamics into the usual 
$x$-based circular templates 
that require the 2PN-accurate reactive 
dynamics, given by Eqs.~(\ref{Eq:Phasing_x}).
The circular TaylorEt approximant 
is again invoked to model our fiducial GW
signal mainly to point out that the FFs are not improved for binaries having initial $e_t < 0.001$ .   

 The performances of the TaylorEt circular templates against our 
fiducial GW signals from eccentric binaries are compared in Table~\ref{Table:XC_Et}.
The numbers, not surprisingly, demonstrate  that the circular inspiral templates based
on the TaylorEt approximant is both effectual and faithful in capturing
GWs from compact binaries having tiny orbital eccentricities.  
The listed biases in the total masses of the canonical binaries are 
consistent with the fact that both eccentricity and higher values for the total mass
are capable of decreasing the number of GW cycles in a given GW frequency window.

 The numbers displayed in Tables~\ref{Table:XC_T4(2+2)} and \ref{Table:XC_AdT1_CAdT1}
clearly indicate that GW signals from 
inspiralling compact binaries having  tiny orbital eccentricities,  
constructed by adapting the phasing formalism, detailed in Ref.~\cite{DGI,KG06} 
and summarized in Sec.~\ref{Sec_II:Phasing},
can not be captured by the usual $x$-based  adiabatic and complete adiabatic TaylorT1 and 
adiabatic TaylorT4 circular templates that incorporate the 2PN-accurate reactive dynamics.  
This is should be of definite concern for the GW 
community because the expected
GWs from astrophysical compact binaries will have tiny orbital 
eccentricities and the GW data analysts belonging to the LSC employed, even in the recent past,
the traditional $x$-based 2PN order circular inspiral templates to search for GWs from 
non-spinning compact binaries \cite{lsc_grb}.   
The numbers displayed in Tables~\ref{Table:XC_T4(2+2)} and \ref{Table:XC_AdT1_CAdT1}
evince that the dominant reason for the above unexpected  conclusion 
is the inability of the 
$x$-based circular inspiral templates, based on the two TaylorT1 and T4 approximants, 
to capture GW signals modeled after the TaylorEt approximant, in an effectual and
faithful manner.

   From a purely mathematical point of view, it may be argued that
the circular TaylorEt and $x$-based approximants should be identical as 
they differ only in the use of the PN-expansion parameter.
This implies that the observed differences in GW phase evolutions
under various PN approximants may be attributed to the 
uncontrolled higher order terms.
However, the present analysis and the one detailed in Ref.~\cite{BGT08}
indicate that 
for the data analysis considerations the inspiral templates 
belonging to the TaylorEt and the $x$-based approximants should  be 
treated to be different. 
From a physical point of view, these approximants are also different due to the 
following reasons.
	  It should be noted that due to 
	 the use of the standard energy balance argument, various circular
	 inspiral templates
	 model GWs from compact binaries evolving along a sequence of circular orbits under the
	 action of gravitational radiation reaction.
For the $x$-based templates, the above circular
orbits can be treated to be Newtonian-accurate due to the use of
Eq.~(\ref{Eq:Phas_x_dphdt_1}) for $d \phi/dt$.
However, the Taylor Et approximant requires the 3PN-accurate expression for $d \phi/dt$
and therefore the sequence circular orbits can be considered to be 3PN-accurate.
The differences in the above mentioned 
circular orbits imply that the TaylorEt approximant is specified by
two separate PN-accurate differential equations, while
the TaylorT1 and TaylorT4 approximants are governed by
just one PN-accurate differential equation.
Further and as noted earlier, 
for a given GW frequency window, 
the limits of the integration for
the $x$-based approximants,
irrespective of whether one is
interested in Newtonian or 3.5PN accurate templates,
are the same and are independent of $\eta$.
However, the limits of the integration for the TaylorEt approximant in the present study
crucially depend on the underlying 3PN-accurate conservative dynamics and
do depend on $\eta$.

 It may be argued that if one constructs eccentric GW signals using 
the generalized $x$ parameter, introduced in an ad-hoc manner in Refs.~\cite{ABIQ,Hinder08},
and $e_t$, 
the 2PN order TaylorT4 approximant (and hence LSC's 2PN-accurate TaylorF2 approximant) 
will have no difficulty in capturing such 
eccentric GW signals.  
This is because the TaylorT4 approximant indeed provides the circular limit of
GW phasing for eccentric binaries that employ the generalized $x$ parameter and $e_t$.
Let us (again) recall that the use of $\omega \equiv d \phi/dt$ as a PN expansion 
parameter for the circular inspiral is justified due to the fact that it provides
the instantaneous orbital angular frequency. Further, the emitted 
instantaneous GW angular frequency
is related to its orbital counterpart by a factor of two in the case the restricted 
PN-accurate circular templates.
However, as noted in the previous section and clearly visible in Eqs.~(A10)-(A14) in Ref.~\cite{Hinder08}, 
the PN expansion parameter $\omega \equiv <d \phi/dt>$ present in 
the generalized $x$ parameter \emph {will not} provide 
the instantaneous orbital angular frequency in the case of 
compact binaries in inspiralling eccentric orbits. Further, 
the GW spectrum associated with compact binaries in PN-accurate eccentric orbits 
\emph { is not} 
expressible as integer multiples of $<d \phi/dt> = n (1 +k) $ \cite{TG07}.
Therefore, it is not clear why one should  employ $\omega \equiv <d \phi/dt>$
as the PN expansion parameter to perform GW phasing for eccentric binaries, when
there exists a more primary and gauge-invariant variable like $\xi$ that is related to 
the binary's instantaneous orbital energy. 

  The practitioners and proponents of the $x$-based inspiral $h_{\times,+}(t)$, both circular and eccentric,
may point out that the PN-accurate GW phasing for eccentric binaries in terms of 
the generalized $x$ parameter and $e_t$ are closer to what is provided by NR,
as reported in Ref.~\cite{Hinder08}.
In our opinion, this is most likely due to the ability of the circular 2PN-accurate TaylorT4 approximant 
to track fairly closely the GW phase evolution, 
originating from NR, associated with the equal-mass binary black hole inspirals that last several orbits 
near their respective last stable orbits\cite{CC07,GHHB}. 
An inspection of Fig.~4 in Ref.~\cite{GHHB} reveals that at the 2PN reactive order
the accumulated GW differences between TaylorEt-TaylorT1  and TaylorEt-TaylorT4 are
$\sim 4.4$ radians and $ \sim 2.6$ radians, respectively.
These differences in the accumulated GW phase   
may be attributed to the uncontrolled higher order terms present in the various Taylor approximants
and are partially responsible for our FF results.
The statement about the partial responsibility is due to our observation that
both the TaylorT1 and T4 approximants are fairly effectual in capturing 
GWs from equal mass circular 
compact binaries, based on the TaylorEt approximant and having $m \geq 30\,M_{\odot}$.
However, these $x$-based approximants are not effectual (and faithful) in the case of 
low mass binaries.
We note that similar results prevail even when the RR order is  
increased to the 3.5PN relative  order while dealing with equal-mass circular inspirals\cite{BGT08}.
This is despite the observation 
that the GW phase evolution under the circular TaylorEt approximant at 3.5PN order is 
fairly close to NR (and hence to the TaylorT4 approximant at 3.5PN order) 
in comparison with what is provided by
the TaylorEt approximant at the 2PN reactive  order \cite{GHHB}.

 The Fig.~4 in Ref.~\cite{GHHB}
indicates that the PN-accurate GW phase evolutions
prescribed by the circular TaylorEt and TaylorT4 approximants at 2.5PN reactive order
can not be substantially different from each other.
Therefore, it is natural to expect that these two approximants will be
effectual and faithful to each other for data analysis considerations.
The entries listed in
{ Table~\ref{Table:Et(3+2.5)_T4(2+2.5)}}
support somewhat different conclusions,
similar to what is detailed in Ref.~\cite{BGT08}. 
In other words, for low $m$ compact binaries, the TaylorT4 approximant is not effectual 
and in the case of high $m$ binaries, the $x$-based approximant 
is effectual at the expense of inviting some what large biases in $\eta$.

  The earlier discussions that supported the use of $\xi$ as an appropriate
PN expansion parameter for doing GW phasing for eccentric binaries
and the monotonic convergence to the NR based GW phase evolution,
exhibited by its circular counterpart, suggest that 
the TaylorEt approximant at 3.5PN order is
a promising candidate to search for inspiral GWs from
astrophysical compact binaries that are surely going to have
tiny orbital eccentricities.
Further, the Fig.4 in Ref.~\cite{GHHB} reveals that the 
accumulated phase disagreement between the 3.5PN order TaylorEt approximant
and 2PN order TaylorT4 approximant is $\sim 1.00$ radians while 
considering an equal-mass binary inspiral that last around nine orbits during
its late inspiral stage. Therefore, we probed how effectual and faithful
are these two distinct approximants and the results are listed in 
{ Table~\ref{Table:Et(3+3.5)_T4(2RR)}.}
We observe that FFs that are greater than $0.97$ are only possible for binaries that have
$m >35 \,M_{\odot}$ and for other stellar mass binaries, the 2PN order TaylorT4 
templates are fairly ineffectual. 
It is worthwhile to note that this experiment is also 
prompted by the observation that the LSC employed 2PN-accurate TaylorF2 
templates that are identical to the 2PN-accurate TaylorT4 approximant 
for data analysis considerations \cite{lsc_grb}.

 It is plausible that the inability of the 2PN-accurate TaylorT4 approximant to capture
inspiral GWs modeled after the TaylorEt approximant or its eccentric extension 
can have astrophysical implications.
Recently, the LSC reported that it is rather impossible to have a compact  
binary progenitor for the gamma-ray-burst GRB~070201 having
its masses in the range $1 M_{\odot} < m_1 < 3\, M_{\odot} $
and $1 M_{\odot} < m_2 < 40\, M_{\odot} $
 to be located in the spiral arm of the Andromeda galaxy \cite{lsc_grb}.
This conclusion was reached by analyzing the LIGO data, associated with the science run ${\rm S}5$,
lasting around 180 seconds around the time of the GRB~070201.
Interestingly, the LSC employed the 2PN-accurate TaylorF2 templates 
to filter the LIGO data while searching for 
inspiral GWs from non-spinning compact binaries in the above mass range \cite{PB_pc}.
It is astrophysically quite  possible that the above progenitor candidate
will have some tiny non-zero residual orbital eccentricity and
therefore, we explored the ability of the 2PN-accurate TaylorT4 approximant to
extract the inspiral GWs from compact binaries having residual initial $e_t \sim 10^{-3}$.
The use of the 2PN-accurate TaylorT4 approximant is justified as 
the 2PN-accurate TaylorF2 templates are only the analytical Fourier-domain version of 
the time-domain TaylorT4 templates. 
The numbers listed in
{ Table~\ref{Table:GRB_XC_2PN_T4}}
indicate that the TaylorT4 approximant is 
effectual with respect to our eccentric GW signal only in the high total mass scenario like
the case involving $m_1= 3\,M_{\odot}\,,\,\, m_2= 40\,M_{\odot}$ and in these cases
the search templates are highly unfaithful.
For other mass combinations, the TaylorT4 templates are neither effectual nor 
faithful.  The entries listed in 
{ Table~\ref{Table:GRB_XC_2PN_T4}}
force us to state that further 
investigations involving the relevant LIGO data sets and the various PN-accurate LAL templates
would be very useful to clarify the real implications of our FF results.
Finally, we would like to state that 
the above detailed FF experiments 
were repeated several times 
by employing a wide range of initial 
values for $m$ and $\eta$, required by the {\sl amoeba} routine,
and the numbers listed in our tables  belong to the 
runs that provided the best possible FFs.

 The results arising from our various numerical experiments force us to conclude that
it is rather desirable (and may be urgent)
to explore the ability of various $x$-based PN-accurate inspiral templates, prescribed in
Refs.~\cite{DIS01,CC07},
to capture inspiral GW signals from astrophysical compact binaries that are bound to have 
tiny residual orbital eccentricities.

%
\section{Conclusions} 
\label{sec:SIV}

  In this paper, we probed the 
ability of a number of traditional PN-accurate 
circular inspiral templates,
based on the adiabatic TaylorT1, completely adiabatic TaylorT1 and adiabatic TaylorT4 approximants,
to capture
inspiral GWs from compact binaries having tiny residual orbital eccentricities.
Our inspiral GW signals are modeled by adapting the phasing formalism of Ref.~\cite{DGI} 
and therefore originate from 
compact binaries moving in 3PN-accurate 
eccentric orbits, specified by $\xi$ and $e_t$,
that are perturbed by the 2PN-accurate RR effects. 
We demonstrate that the traditional circular templates that incorporate 2PN-accurate RR effects 
are neither effectual nor faithful in capturing our GW signals having tiny 
orbital eccentricities. 
However, PN-accurate circular templates based on the recently introduced TaylorEt approximant 
are both effectual and faithful with respect to GW signals having small orbital eccentricities like $0.1$.
We explained the physical motivation for employing $\xi$, a measure of the 
instantaneous
orbital binding energy,
to describe PN-accurate circular and eccentric orbits. 
With the help of a number of FF experiments, we explored the implications of using only the 
traditional PN-accurate inspiral templates to filter the LIGO data.

  We are planning and pursuing a number investigations 
motivated by the physical, numerical relativity and data analysis considerations 
to probe various aspects of the TaylorEt approximant, both circular and eccentric.
It will be desirable to extend the present study by including at least the 3PN-accurate 
RR effects in an accurate and efficient manner
and which include the $\tilde c_{\alpha}  $ contributions while describing the inspiral dynamics
of PN-accurate eccentric binaries, relevant for both the ground-based and the proposed space-based 
GW interferometers.
Using such a PN-accurate dynamics, it will be interesting to
reexamine what is pursued in Ref.~\cite{Hinder08},
while employing $\xi$ rather than $\omega \equiv <\frac{d \phi}{dt}>$ to describe the PN-accurate orbit.
We are in the middle of incorporating the spin effects into the present study 
with the help of inputs that 
generalize what is 
detailed in Ref.~\cite{HHBG}.

 \acknowledgments
It is our pleasure to thank 
Gerhard Sch\"afer for helpful discussions and persistent
encouragement.
This work is supported in part by
DFG's SFB/TR 7 ``Gravitational Wave Astronomy'' 
and DLR (Deutsches Zentrum f\"ur Luft- und Raumfahrt).


\begin{widetext}

\begin{figure}[ht]
\includegraphics[height=10cm, width=0.9\textwidth, angle=0]{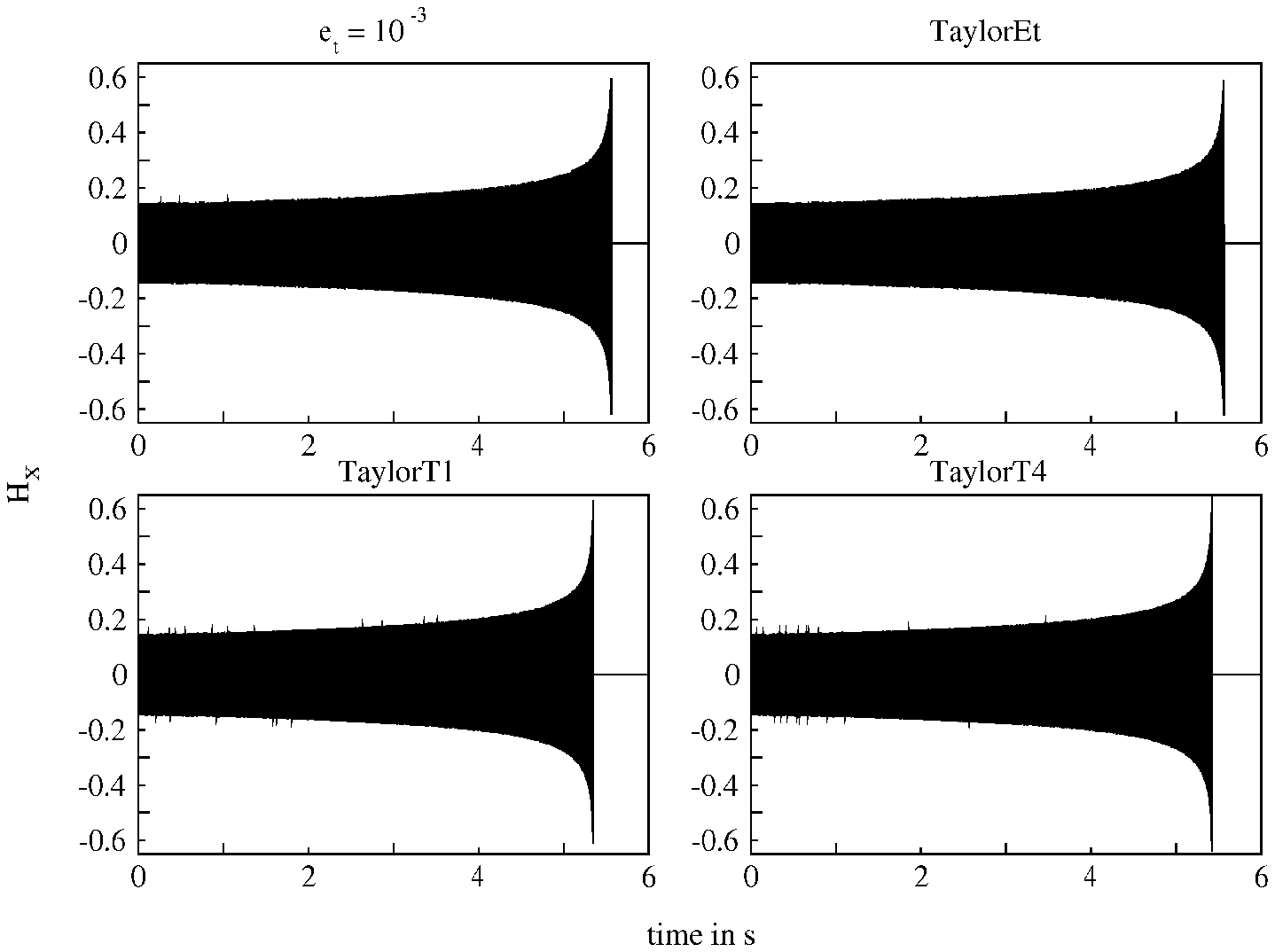}\\[0.1cm]
\caption{
\label{FIG:h_c_e_T1_T4_Et}
A set of plots that display the temporal evolution for 
 $H_{\times}(t)=h_\times\big|_{\rm Q}(t) / \left( \frac{G m \eta\, C}{R' c^2}\right)$, originating from a 
neutron star-black hole binary having $m_1= 10\,M_{\odot}$ and $ m_2 = 1.4\,M_{\odot}$,
and in the initial LIGO frequency window.
The top left panel depicts an eccentric orbital evolution of  Sec.~\ref{Sec_II:Phasing}, having an initial orbital 
$e_t= 10^{-3}$. The other three plots are for the circular templates modeled after the TaylorEt 
approximant, the adiabatic TaylorT1 and T4 approximants and these a given by 
Eqs.~(\ref{Eq:P_Et}), (\ref{Eq:P_adT1}) and (\ref{Eq:P_T4}), respectively.
The accumulated number of GW cycles in the initial LIGO frequency window, defined by 
40Hz-$\left( 6^{3/2}\, G\, m\, \pi /c^3 \right )^{-1 } $ Hz,
 are $\sim 346.264, 346.257, 326.048, 333.622$ for the eccentric inspiral 
and circular inspirals defined by the TaylorEt, adiabatic TaylorT1 and T4 approximants,
respectively.     
}
\centering
\end{figure}

\newpage

\begin{table}
\caption{
{
The initial LIGO FFs involving
the 2PN-accurate circular TaylorT4 approximant  
for the three types of 
canonical binaries.
The fiducial GW signals from compact binaries  
in inspiralling eccentric orbits are constructed using the phasing formalism
detailed in Sec.~\ref{Sec_II:Phasing}.
The templates that provide the listed fully maximized
minimax-overlap (FF) are characterized by $m_t$ and $\eta_t$. 
In the case of double neutron-star binaries, we terminate the the temporal evolution of both the GW
signal and circular templates when their respective GW
frequencies reach 1000Hz. 
In the $e_t \equiv 0$ cases, we employ the TaylorEt approximant to model the fiducial GW signals 
and the associated numbers can be considered to represent what to expect for initial $e_t < 10^{-3}$.
The numbers indicate that the 2PN-accurate TaylorT4 templates are neither
effectual nor faithful while dealing with inspiral GWs from compact binaries having
tiny orbital eccentricities. 
}
}
\label{Table:XC_T4(2+2)}
\begin{tabular}{||l r|r|r|r|r|}
\hline
$m_1/ M_{\odot} : m_2/ M_{\odot}$ &	& $1.4 : 1.4$ 	& $1.4 : 10.0$ & $10.0 : 10.0$ \\
\hline
$e_t\equiv0 $&			FF		& 0.822 & 0.939 & 0.957 \\
$	    $&			$m_t $		& 2.800 & 6.906 & 19.04 \\
$	    $&			$\eta_t$	& 0.248 & 0.250 & 0.247 \\
\hline
$e_t=0.001  $&			FF		& 0.821 & 0.939 & 0.951 \\
$	    $&			$m_t $		& 2.800 & 6.906 & 20.00 \\
$	    $&			$\eta_t$	& 0.248 & 0.250 & 0.224 \\
\hline
$e_t=0.01$&			FF		& 0.821 & 0.938 & 0.951 \\
$	    $&			$m_t $		& 2.800 & 6.909 & 20.00 \\
$	    $&			$\eta_t$	& 0.248 & 0.250 & 0.225 \\
\hline
$e_t=0.10$&			FF		& 0.708 & 0.830 & 0.830 \\
$	    $&			$m_t$		& 2.800 & 6.943 & 19.00 \\
$	    $&			$\eta_t$	& 0.250 & 0.250 & 0.250 \\
\hline
\end{tabular}
\end{table}


\begin{table}
\caption{
{ Values of the FFs and the associated biases in $m$ and $\eta$ values, while employing
the circular inspiral templates based on the adiabatic T1 and the complete adiabatic
T1 approximants. 
The complete adiabatic TaylorT1 templates are employed to incorporate
the effects of 3PN-accurate conservative dynamics into the $x$-based circular templates \cite{AIRS}.
The conclusions and other details are similar to what is observed in Table
\ref{Table:XC_T4(2+2)}.
}
}
\label{Table:XC_AdT1_CAdT1}
\begin{tabular}{||l r|r|r|r|r|}
\hline
$m_1/ M_{\odot} : m_2/ M_{\odot}$ &	& $1.4 : 1.4$ 	& $1.4 : 10.0$ & $10.0 : 10.0$ \\
\hline
\multicolumn{5}{||c|}{ Standard adiabatic TaylorT1 templates 	} \\
\hline
$e_t\equiv0  $&			FF	& 0.771	& 0.843	& 0.894  \\
$	$&			$m_t $		& 2.785	& 6.845 & 18.55  \\
$	$&			$\eta_t$	& 0.250	& 0.250 & 0.243  \\
\hline
$e_t=0.001$&			FF		& 0.768 & 0.842 & 0.891  \\ 
$	$&			$m_t $		& 2.800 & 6.846 & 18.57  \\ 
$	$&			$\eta_t$	& 0.248 & 0.250 & 0.242  \\ 
\hline
$e_t=0.01$&			FF		& 0.766 & 0.841 & 0.841  \\ 
$	 $&			$m_t $		& 2.800 & 6.834 & 6.837  \\ 
$	 $&			$\eta_t$	& 0.248 & 0.250 & 0.250  \\ 
\hline
$e_t=0.10$&			FF		& 0.677 & 0.766 & 0.778  \\ 
$	 $&			$m_t$		& 2.800 & 6.876 & 20.01  \\ 
$	 $&			$\eta_t$	& 0.249 & 0.250 & 0.216  \\ 
\hline
\multicolumn{5}{||c|}{ Complete adiabatic TaylorT1 templates	} \\
\hline

$e_t\equiv0$& 			FF		& 0.761	 & 0.800 & 0.878 \\
$	$&			$m_t $		& 2.783	 & 6.820 & 18.41 \\
$	$&			$\eta_t$	& 0.250	 & 0.250 & 0.245 \\
\hline
$e_t=0.001$&			FF		& 0.761 & 0.800 & 0.878 \\ 
$	$&			$m_t $		& 2.786 & 6.821 & 18.41 \\ 
$	$&			$\eta_t$	& 0.250 & 0.250 & 0.246 \\ 
\hline
$e_t=0.01$&			FF		& 0.758 & 0.800 & 0.878 \\ 
$	 $&			$m_t $		& 2.800 & 6.820 & 18.41 \\ 
$	 $&			$\eta_t$	& 0.247 & 0.250 & 0.247 \\ 
\hline
$e_t=0.10$&			FF		& 0.673 & 0.754 & 0.838 \\ 
$	 $&			$m_t$		& 2.800 & 6.867 & 18.55 \\ 
$	 $&			$\eta_t$	& 0.249 & 0.249 & 0.247 \\ 
\hline
\end{tabular}
\end{table}


\begin{table}
\caption{
{The results of FF computations that employ the circular 
templates based on the TaylorEt approximant. 
It is important to note that the temporal evolution for both 
the eccentric GW signals and the circular templates require the 3PN accurate
conservative and the 2PN-accurate reactive dynamics
The rest of the details are as
in Tables \ref{Table:XC_T4(2+2)} and \ref{Table:XC_AdT1_CAdT1}.
The displayed values clearly demonstrate, not surprisingly, that the circular
TaylorEt templates are both effectual and faithful with respect to our inspiral
GW signals having tiny orbital eccentricities.}
}
\label{Table:XC_Et}
\begin{tabular}{||l r|r|r|r|r|}
\hline
$m_1/ M_{\odot} : m_2/ M_{\odot}$ &	& $1.4 : 1.4$ 	& $1.4 : 10.0$ & $10.0 : 10.0$ \\
\hline
$e_t=0.001$& \hspace{0.2\textwidth}FF		& 0.999 & 1.000 & 1.000\\
$	$&			$m_t $		& 2.824 & 11.50 & 20.17\\
$	$&			$\eta_t$	& 0.246 & 0.106 & 0.246\\
\hline
$e_t=0.010$&			FF		& 0.999 & 1.000 & 1.000\\
$	 $&			$m_t $		& 2.823 & 11.45 & 20.18\\
$	 $&			$\eta_t$	& 0.247 & 0.107 & 0.246\\
\hline
$e_t=0.100$&			FF		& 0.804 & 0.921 & 0.956\\
$	 $&			$m_t$		& 2.811 & 11.44 & 20.20\\
$	 $&			$\eta_t$	& 0.250 & 0.108 & 0.249\\
\hline
\end{tabular}
\end{table}


\begin{table}
\caption{
{The initial LIGO FFs involving GW signals modeled after the circular TaylorEt approximant 
and the circular templates based on the TaylorT4 approximant, both requiring 
2.5PN-accurate reactive dynamics. 
The TaylorEt GW signals are constructed using the 3PN-accurate expression for $d \phi/dt$ 
and the 2.5PN-accurate expression for $d \xi/dt$, while the the  TaylorT4 approximant
requires the 2.5PN-accurate expression for $dx/dt$. These PN-accurate expressions are 
extractable from Eqs.~(6) and (7) in Ref.~\cite{BGT08}.
We conclude that for low mass binaries, the TaylorT4 approximant is not very effectual 
and in the case of high $m$ binaries, the circular templates are fairly unfaithful. 
}
}
\label{Table:Et(3+2.5)_T4(2+2.5)}
\begin{tabular}{||l r|r|r|}
\hline
$m_1:m_2$       & FF    & $m_t$ & $\eta_t $ \\
\hline
 5-5            & 0.933 & 9.900 & 0.250 \\
 3-9            & 0.868 & 9.888 & 0.248 \\
 3-12           & 0.897 & 11.26 & 0.249 \\
 5-10:          & 0.903 & 13.47 & 0.250 \\
 5-15:          & 0.937 & 16.20 & 0.249 \\
 10-10          & 0.995 & 20.51 & 0.232 \\
 15-15          & 0.995 & 30.72 & 0.231 \\
\hline
\end{tabular}
\end{table}


\begin{table}
\caption{
{ The results originating from one of our FF experiments motivated by the Fig.~4 in Ref.~\cite{GHHB}.
The inspiral GW signals are based on the circular TaylorEt approximant, given 
by Eq.~(7) in Ref.~\cite{BGT08} and therefore specified by
the 3PN-accurate expression for $d \phi/dt$ and the 3.5PN-accurate expression for $d \xi/dt$. The 
circular templates are given by the 2PN-accurate TaylorT4 approximant. 
Somewhat higher FFs are observed only for high mass binaries with $m \ge 30\,M_{\odot}$. 
 }
}
\label{Table:Et(3+3.5)_T4(2RR)}
\begin{tabular}{||l |r|r|r|}
\hline
$m_1/ M_{\odot} : m_2/ M_{\odot}$
		& FF   & $m_t$     & $\eta_t$ \\ 
\hline
5-5		& 0.849 &  9.744 & 0.247 \\
1.4-10		& 0.877 &  6.891 & 0.250 \\
3-9		& 0.868 &  9.871 & 0.248 \\
3-12		& 0.858 & 14.058 & 0.163 \\
10-10		& 0.942 & 18.940 & 0.247 \\
15-15		& 0.971 & 27.701 & 0.241 \\

\hline
\end{tabular}
\end{table}


\begin{table}
\caption{
{ The initial LIGO FF values in the mass range  relevant for the 
GRB~070201 progenitor following Ref.~\cite{lsc_grb}. 
The fiducial inspiral GW signals are from compact binaries having 
initial $e_t = 10^{-3}$ and computed using the phasing formalism of Sec.~\ref{Sec_II:Phasing}.
The value of $e_t$ can be treated to be representative of still lower residual initial 
eccentricities and
the circular templates are based on the 2PN-accurate TaylorT4 approximant.
Even though, the employed circular templates are neither effectual nor faithful for 
several mass combinations,
we reiterate that
further investigations involving the relevant LIGO data should be useful to 
make any physical implications of our ideal FF experiments.
 }
}
\label{Table:GRB_XC_2PN_T4}
\begin{tabular}{||l |r|r|r|}
\hline
$m_1/ M_{\odot} : m_2/ M_{\odot}$
		& FF   & $m_t$     & $\eta_{t}$ \\ 
\hline
3- 1		& 0.837	&  3.363 & 0.250 \\
1-10		& 0.909	&  5.702 & 0.250 \\
3-10		& 0.876	& 10.335 & 0.249 \\
1-20		& 0.937	&  7.634 & 0.250 \\
3-20		& 0.945 & 14.029 & 0.250 \\
1-40		& 0.966 & 11.680 & 0.195 \\
3-40		& 0.972	& 20.675 & 0.208 \\

\hline
\end{tabular}
\end{table}

\end{widetext}

\end{document}